 \documentclass[preprint2]{aastex}

\usepackage{epsfig}
\usepackage{epstopdf}

\usepackage{ulem}

\shorttitle{}
\shortauthors{Godet et al.}

\begin{document}


\title{Implications of the delayed 2013 outburst of the ESO 243-49 HLX-1}


\author{O. Godet$^{1,2}$}
\affil{$^1$ Institut de Recherche en Astrophysique and Plan\'etologie (IRAP), Universit\'e de Toulouse, UPS, 9 Avenue du colonel Roche, 31028 Toulouse Cedex 4, France\\
 $^2$ CNRS, UMR5277, 31028 Toulouse, France}
\author{J. C. Lombardi$^3$}
\affil{$^3$ Department of Physics, Allegheny College, Meadville, PA 16335, USA}
\author{F. Antonini$^4$}
\affil{$^4$ Canadian Institute for Theoretical Astrophysics, University of
  Toronto, 60 St. George Street, Toronto, Ontario M5S 3H8, Canada}
\author{D. Barret, N. A. Webb$^{1,2}$}
\and
\author{J. Vingless$^3$, M. Thomas$^3$}

\begin{abstract}

After showing four quasi-periodic outbursts spaced by $\sim 1$ year from 2009
to 2012, the hyper luminous X-ray source ESO\,243-49 HLX-1, currently the best
intermediate mass black hole (IMBH) candidate, showed an outburst in 2013
delayed by more than a month.  In Lasota et al. (2011),
we proposed that the X-ray lightcurve is the result of enhanced mass transfer
episodes at periapsis from a donor star orbiting the IMBH in a highly
eccentric orbit. In this scenario, the delay can be explained only if the
orbital parameters can change suddenly from orbit to orbit.  To investigate
this, we ran Newtonian smooth particle hydrodynamical (SPH) simulations
starting with an incoming donor approaching an IMBH on a parabolic orbit. We
survey a large parameter space by varying the star-to-BH mass ratio
($10^{-5}-10^{-3}$) and the periapsis separation $r_p$ from 2.2 to $2.7~r_t$
with $r_t$, the tidal radius. To model the donor, we choose several polytropes
($\Gamma = 5/2,~n=3/2$, $\Gamma=3/2,~n=2$, $\Gamma=5/3,~n=2$ and
$\Gamma=5/3,~n=3$).
 
Once the system is formed, the orbital period decreases until reaching a
minimum that may be shallow. Then, the period tends to increase over several
periapsis passages due to tidal effects and increasing mass transfer, leading
ultimately to the ejection of the donor. We show that the development of
stochastic fluctuations inside the donor by adding or removing orbital energy
from the system could lead to sudden changes in the orbital period from orbit
to orbit with the appropriate order of magnitude of what has been observed for
HLX-1.  We also show that given the constraints on the BH mass ($M_{\rm BH} >
10^4~M_\odot$) and assuming that the HLX-1 system is currently near a minimum
in period of $\sim 1$ yr, the donor has to be a white dwarf or a stripped
giant core. We predict that if HLX-1 is indeed emerging from a minimum in
orbital period, then the period would generally increase with each passage,
although substantial stochastic fluctuations can be superposed on this trend.

\end{abstract}


\section{Introduction}

Stellar mass black holes (BHs) are known to be the remnants of stellar
activity with masses typically ranging from $\sim 3-20$ M$_{\odot}$. At the
other end, supermassive BHs are thought to be present in the core of most
massive galaxies with masses typically ranging from $\sim 10^{6-10}$
M$_{\odot}$. However, the way supermassive BHs formed is still poorly
understood. Most theories agree that supermassive BHs we observe today have
been formed from lighter BH seeds present in the early Universe. What differs
between models is the growth mechanisms and the mass range of the BH seeds
(e.g. Alexander \& Hickox 2012). Super-Eddington accretion of matter onto
massive stellar mass BHs of $\sim 100$ solar masses has been invoked to
explain the growth of supermassive BHs (e.g. Kawaguchi et al. 2004). Other
scenarios propose the mergers of BHs with masses $\sim 10^2$ to $\sim 10^5$
solar masses, the so-called intermediate mass BHs (IMBHs -- Madau \& Rees
2001). In some models, both mechanisms are expected to take place. In the
IMBHs scenarios, a fundamental question is then to understand how to form and
how such BHs evolve (Miller \& Colbert 2004).  Despite long and thorough
searches over the past last decades in different astrophysical objects such as
globular clusters (Pooley \& Rappaport 2006; Strader et al. 2012) and the
low-mass tail of AGN (e.g. Greene \& Ho 2004), the observational evidence for
their existence is weak.

The serendipitous discovery of 2XMM J011028.1--460421 (hereafter Hyper
Luminous X-ray source - HLX-1) with XMM-Newton on 23 November 2004 in the
outskirts of the edge-on spiral galaxy ESO 243--49 at a redshift of 0.0224
marked a milestone with the most secure identification of an IMBH (Farrell et
al. 2009 -- hereafter F09). Wiersema et al. (2010) confirmed the association
of HLX-1 to ESO 243-49 (see also Soria et al. 2013). With a 0.2-10 keV
unabsorbed luminosity reaching $1.3\times 10^{42}$ erg s$^{-1}$ at peak, HLX-1
is the brightest Ultra Luminous X-ray source (ULX -- Roberts 2007; Feng \&
Soria 2011) detected so far. Spectral modeling of X-ray data with
sophisticated accretion disk models (Davis et al. 2011; Godet et al. 2012;
Straub et al. 2014) and Eddington scaling of X-ray data (Servillat et
al. 2011) gave us a range of mass estimates around $\sim
10^4~\mathrm{M}_\odot$ and recent radio data provide an upper limit on the BH
mass of $9\times 10^4~\mathrm{M}_\odot$ assuming Eddington scaling (Webb et
al. 2012). Multi-wavelength observations of HLX-1 over the past five years
enabled us to show that HLX-1 displays several properties similar to those
observed in stellar mass BH X-ray binaries (e.g. Remillard \& McClintock
2006): i) regular outbursts with state transitions, but with X-ray
luminosities orders of magnitude larger (Godet et al. 2009b; Servillat et
al. 2011); ii) the first ever detection of radio transient emission in a ULX
interpreted as discrete jet ejection events following the hard-to-soft
transitions (Webb et al. 2012). In 2012, we ran a series of contemporaneous
optical (VLT) and X-ray ({\it Swift}-XRT) observations over the rise of the
outburst in order to put some constraints on the outburst mechanism. Our data
showed the optical may rise just before the X-rays, the delay being shorter
than a day (Webb et al. 2014). 

\begin{figure*}
\begin{center}
\includegraphics[angle=0,scale=0.8]{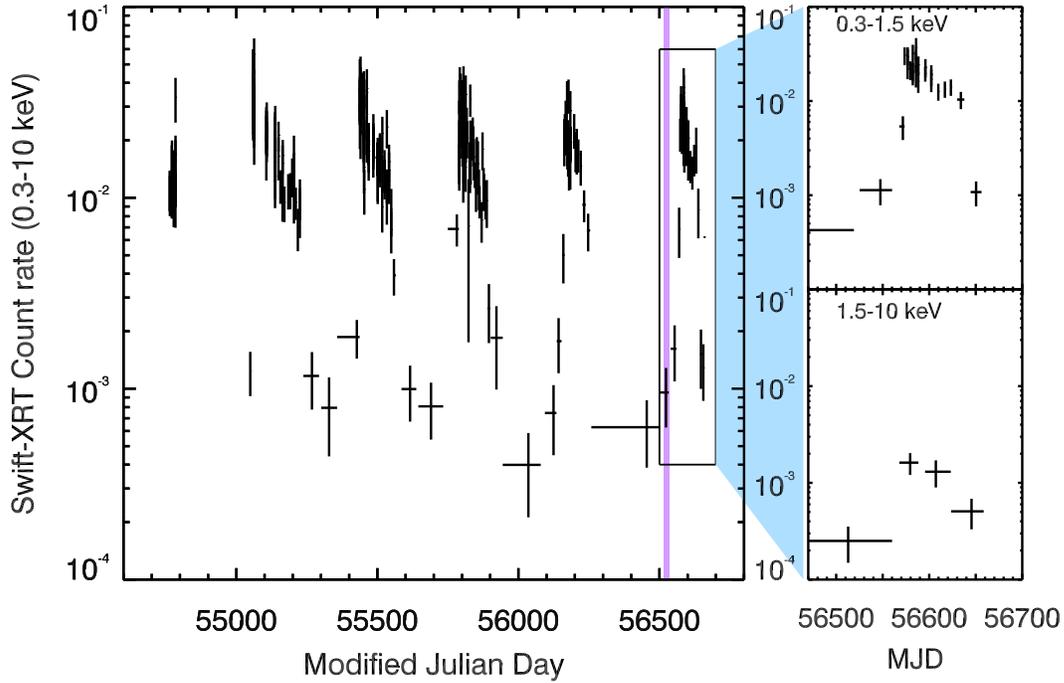}
\caption{{\it Swift}-XRT Photon Counting lightcurve of HLX-1 in the 0.3-10 keV
  band from 24$^\mathrm{th}$ October 2008 to 31$^\mathrm{st}$ December
  2013. The two panels on the right show a close-up of the X-ray lightcurve
  for the outburst in 2013: (top) in the 0.3-1.5 keV range; (bottom) in the
  1.5-10 keV range. Since then, the source is in the low/hard state. The
  lightcurve was produced using the online lightcurve generator made available
  on the {\it Swift} UK website. We used a dynamical binning of 20 counts per
  bin. The vertical colored band corresponds to the time interval when the
  source was expected to enter in outburst in 2013 if it kept the timing
  observed from 2009 to 2012 (between 10$^\mathrm{th}$ and 30$^\mathrm{th}$ of
  August).
\label{fig1}}
\end{center}
\end{figure*}

The {\it Swift}-XRT (Gehrels et al. 2004; Burrows et al. 2005) light-curve
from 2009 to 2013 shows Fast Rise and Exponential Decay (FRED) outbursts
separated by an apparent recurrence time of nearly a year (see
Figure~\ref{fig1}). In Lasota et al. (2011 -- hereafter L11), we investigated
the possibility to explain these outbursts as well as the spectral evolution
during outburst in the framework of the Dwarf Instability Model (DIM --
e.g. Lasota 2001). We demonstrated that outbursts of HLX-1 {\sl cannot} be due
to the thermal-viscous instability given the source distance of 95
Mpc. Instead, we proposed that the X-ray light-curve is the result of enhanced
mass transfer episodes from an evolved (Asymptotic Giant Branch) star orbiting
the IMBH in a highly eccentric orbit when the star passing at periapsis is
tidally stripped. Given the small delay between the optical and X-rays, we
also stressed that the mass delivery radius is rather small implying in the
framework of the L11 scenario that the orbit must be highly eccentric.

Recently, Miller et al. (2014 -- hereafter M14) proposed a mass-transfer
scenario based on wind-fed accretion onto an IMBH when the donor star passes
at periapsis. This scenario shows several similarities with the model we
proposed in L11. To be able to account for a peak accretion rate around
$\dot{M}_{\rm peak}\sim 10^{-4}~M_\odot$ yr$^{-1}$, M14 estimated that the
mass loss needed should be at least one order of magnitude larger than
$\dot{M}_{\rm peak}$. Such a high mass loss is far beyond the expected maximum
mass loss for AGB stars. To circumvent this issue, M14 propose that the AGB
star was first disrupted by the IMBH thus losing its envelope and leaving its
core orbiting the IMBH. The bared core would then emit a strong wind driven by
high radiation pressure. This scenario also presents some caveats that will
have to be investigated in detail: i) even with the loss of its envelope, it
is yet to be demonstrated that the star mass loss could reach $\dot{M}_{\rm
  peak}\sim 10^{-3}~M_\odot$ yr$^{-1}$; ii) How does this scenario account for
the almost simultaneous outburst rise seen in X-rays and optical (Webb et
al. 2014)?; iii) AGB stars are known to produce a lot of dust. So, such a high
mass loss should result in a strong extinction in optical. Our optical data
collected so far does not seem to support this idea (Farrell et al. 2012).

 To be complete on alternative models for HLX-1, we note the recent work
  done by King \& Lasota (2014) who made an analogy of HLX-1 behavior with the
  Galactic source SS433. Again, from this work it appears that a precise model
  fine-tuning needs to be done in order to account for some of the HLX-1
  observational properties. We note that the model is unable to explain fully
  some of the most remarkable properties of HLX-1 (e.g. the spectral
  variation). This piece of work again emphasizes how unique is the HLX-1
  system. However, a detailed analysis of this scenario will have to be done
  to show if it is viable or not.

In this paper, we report on data collected in 2013 showing a significant
change in the HLX-1 behavior. Indeed, the X-ray outburst is delayed by more
than a month compared to previous years (see Fig.~\ref{fig1}). This
corresponds to a recurrence time of more than 400 days since the last outburst
peaking on 23$^\mathrm{rd}$ August 2012.

The aim of the present paper is to improve our understanding of the HLX-1
nature using an alternative approach through Smoothed Particle Hydrodynamical
(SPH) simulations. The present paper will focus on the eccentric binary
scenario we proposed in L11. In a forthcoming paper, we will investigate in
detail the M14 scenario. The paper is organized as follows: in Section 2, we
describe the data reduction of the {\it Swift}-XRT data. In Section 3, we
discuss the main features of the X-ray light-curve, especially the apparent
annual decrease in the outburst duration and the average peak count rate from
2009 to 2012. In Section 4, we present the data collected in 2013 and we
compare the results with those obtained in previous years. In Section 5, we
discuss the constraints that the data collected so far put on the eccentric
binary scenario. In Section 6, we present the SPH simulations we run in order
to investigate the possible reasons of the outburst delay, the lifetime of
such a system and the nature of the donor. We also discuss the role of
stochastic fluctuations developing inside the donor star on the orbital period
and the mass loss stripping as well as the role of artificial viscosity on our
results.  Section 7 is devoted to the concluding remarks.

\section{Data analysis}

The {\it Swift}-XRT Photon Counting data (ObsID 31287 \& 32577), unless
otherwise mentioned, were processed using the HEASOFT v6.14, the tool
{\scriptsize XRTPIPELINE} v0.12.8\,\footnote{See {\scriptsize
    http://heasarc.gsfc.nasa.gov/docs/swift/analysis/}} and the calibration
files ({\scriptsize CALDB} version 4.1). We used the grade 0-12 events, giving
slightly higher effective area at higher energies than the grade 0 events, and
a 20 pixel (47.2 arcseconds) radius circle to extract the source and
background spectra using {\scriptsize XSELECT} v2.4c. The background
extraction region was chosen to be close to the source extraction region and
in a region where we are sure that there are no sources present in the
XMM-Newton field of view. The ancillary response files were created using
{\scriptsize XRTMKARF} v0.6.0 and exposure maps generated by {\scriptsize
  XRTEXPOMAP} v0.2.7. We fitted the spectra within {\scriptsize XSPEC} v12.8.1
(Arnaud et al. 1994) using the response file {\scriptsize
  SWXPC0TO12S6$_{-}$20010101V013.RMF}. All the errors quoted below are given
at a 90\% confidence level for one parameter of interest ({\it i.e.}
$\Delta\chi^2=2.706$).  The source redshift being 0.0224 (Wiersema et
al. 2010), we adopted a source distance of $d_L=95$ Mpc using the cosmological
parameters from the WMAP5 results ($H_0 = 71$ km s$^{-1}$ Mpc$^{-1}$,
$\Omega_M = 0.27$ \& $\Omega_\Lambda=0.73$). The X-ray lightcurve was
generated using the online lightcurve generator made available on the {\it
  Swift} UK website (Evans et al. 2009) and using a dynamical count binning of
20 counts per bin.

When the count statistics was enough, the spectra were grouped at a minimum of
20 counts per bin to provide sufficient statistics to use the $\chi^2$
minimization technique. For data in the low flux state, we used 
the C-statistic (Cash 1979) available within {\scriptsize XSPEC} for the
spectral fit.

\section{The X-ray lightcurve}

We set up a monitoring of HLX-1 (PI: O. Godet) with the {\it Swift}-XRT
through the {\it Swift} ToO program over 5 years in order to investigate in
detail the temporal and spectral variability of the source as well as to
trigger our multi-wavelength programs designed to investigate the nature of
the HLX-1 host and how the BH is fed.  Figure~\ref{fig1} shows the X-ray
lightcurve from 24$^\mathrm{th}$ October 2008 to 18$^\mathrm{th}$ December
2013 for a total of 618.7\,ks so far. Below we summarize the main features
seen in the X-ray lightcurve up to the outburst in 2012:
\begin{enumerate}

\item[i)] Before 2012, the lightcurve displays 4 well sampled FRED-like
  outbursts with some re-flare events spaced by an apparent recurrence time of
  nearly a year (L11). The recurrence period appears to be variable (from
  $\sim 350$ to $\sim 370$ days). Since the serendipitous detection of HLX-1
  with XMM-Newton in November 2004, at least 9 outbursts occurred assuming a
  recurrence time of nearly a year. Survey observations with {\it ROSAT} in
  the early nineties gave non-detections (Webb et al. 2010) indicating that
  either the source was in a low-flux state or in quiescence.

\item[ii)] There is a 2-3 weeks plateau phase when the source reaches its
  outburst peak luminosity (L11; Godet et al. 2012).

\item[iii)] When the source is not in outburst, it stays in the low/hard state
  (Godet et al. 2009; Servillat 2011).

\item[iv)] The source switches rapidly (1-2 weeks) to the low/hard state when
  the 0.3-10 keV count rate is less than $\sim 7\times 10^{-3}$ cts s$^{-1}$.

\item[v)] The outburst duration decreases annually passing from $\sim 170$
  days in 2009 to $\sim 93$ days in 2012, while the average peak count rate
  seems to have annually decreased even if it is more noticeable in
  2012. Indeed, the average peak count rate was equal to $0.030\pm 0.002$ cts
  s$^{-1}$ in 2009, whereas it was equal to $0.020\pm 0.002$ cts s$^{-1}$ in
  2012 (90\% confidence level errors).

\end{enumerate}

Our previous works clearly demonstrated that when in outburst the source
emission is dominated by a soft component ($kT \sim 0.17-0.24$ keV)
interpreted as coming from the accretion disk present around the IMBH (Godet
et al. 2009; Servillat et al. 2011; Davis et al. 2011; Godet et al. 2012;
Straub et al. 2014). In such a case, the decrease in the outburst duration and
the average peak count rate implies that less matter is being annually
accreted. In Godet et al. (2012), we stressed that the accreted mass between the
outbursts in 2009 and 2010 has decreased by $\sim 46\%$. This trend continues
for the outbursts in 2011 and 2012.

\section{A delayed outburst}

The observed evolution in the outburst duration and peak count rate may be
related to a gradual decrease in the amount of mass transferred from the
companion star probably leading to a decrease in the total of mass accumulated
in the disk. In order to further investigate this, we proposed a monitoring of
the outburst in 2013 with the {\it Swift}-XRT that started on 10$^\mathrm{th}$
July 2013 and that should end at the end of 2013. Based on a recurrence period
between 350 and 370 days, we expected the outburst to occur between mid-August
and mid-September 2013 at the very latest. However, the X-ray outburst has
been delayed by more than a month, because it started around
2$^\mathrm{nd}$-8$^\mathrm{th}$ October 2013 {\it i.e.}  $\mathrm{MJD} =
565(67-73)$. Until then, the source had remained in a low-flux state (see
Fig.~\ref{fig1}). As shown in Fig.~\ref{fig1}, the outburst in 2013 lasted
$\sim 65-72$\,days. This strengthens the observed trend of the annual decrease
in the outburst duration. The source switched to the low/hard state around
11$^{\rm th}$ December 2013.

We extracted a spectrum prior to the start of the outburst {\it i.e.} from
19$^\mathrm{th}$ April to 28$^\mathrm{th}$ September 2013 (corresponding to a
38\,ks time exposure) and we fitted it using an absorbed powerlaw assuming an
absorption column of $N_H = 4\times 10^{20}$ cm$^{-2}$. We found a good fit
($C-stat/\mathrm{d.o.f.} = 44.3/68$) with a photon index of $\Gamma = 2.5\pm
0.6$ and a normalization of $N_\Gamma = 1.3\pm 0.4 \times 10^{-5}$ ph
keV$^{-1}$ cm$^{-2}$ s$^{-1}$ @ 1 keV. The derived $\Gamma$-value is
consistent with the photon index found previously in the low/hard state
($\Gamma \sim 2-2.2$ -- Servillat et al. 2011; Godet et al. 2012). The
logarithm of the unabsorbed 0.2-10 keV flux is equal to $\log
F_\mathrm{unabs.} = -13.11^{+0.16}_{-0.17}$. This corresponds to a luminosity
of $L_X = 8.4^{+1.3}_{-1.4} \times 10^{40}$ erg cm$^{-2}$ s$^{-1}$. Fixing the
photon index to $2.2$, we found a 0.2-10 keV unabsorbed luminosity of
$7.9^{+1.0}_{-1.3}\times 10^{40}$ erg s$^{-1}$. This X-ray luminosity appears
to be larger than that derived in the low/hard state in 2009 and 2010 (see
Servillat et al. 2011; Godet et al. 2012).

On $\mathrm{MJD} = 56567$ ({\it i.e.} 2$^\mathrm{nd}$ October 2013), we note a
possible rebrightening with a 0.3-10 keV count rate of $9.6 \pm
3.6~(1\,\sigma) \times 10^{-3}$ cts s$^{-1}$. This data point is consistent
with the count rate measured previously in the low/hard state at the
$2\,\sigma$ level. The background subtracted spectrum reveals 5 counts in the
0.3-1.5 keV band and 5 counts in the 1.5-10 keV band.  However, the following
observation performed 3 days later did not show any source indicating that the
flux has probably decreased again. The XRT 1.6\,ks observation performed on
$\mathrm{MJD} = 56573$ ({\it i.e.} 8$^\mathrm{th}$ October 2013) reveals a
brighter X-ray source at the position of HLX-1 with a 0.3-10 keV count rate of
$2.2\pm 0.4 \times 10^{-2}$ cts s$^{-1}$, marking the start of the outburst in
2013. The event on $\mathrm{MJD} = 56567$ may be a precursor. As opposed to
previous years, we note that the peak count rate in 2013 did not decrease
($CR_{\rm peak} = 2.5^{+0.2}_{-0.3}\times 10^{-2}$ cts s$^{-1}$). We extracted
2 background-subtracted spectra from 8$^\mathrm{th}$ October 2013 to
23$^\mathrm{rd}$ October 2013 ($\sim 14$\,ks -- hereafter S1) and from
1$^\mathrm{st}$ October 2013 to 11$^\mathrm{th}$ December 2013 ($\sim 17$\,ks
-- hereafter S2). We fitted S1 using an absorbed {\scriptsize DISKBB} model,
while we used an absorbed {\scriptsize DISKBB + POWERLAW} model to fit
S2. Fixing the absorption column to $4\times 10^{20}$ cm$^{-2}$ and the
powerlaw photon index at the best constrained value of $\Gamma=2.2$ (see
Servillat et al. 2011, Godet et al. 2012), we found a disk temperature of
$0.20\pm0.02$\,keV for S1 and $0.19\pm 0.03$ keV for S2. The 0.2-10 keV
unabsorbed luminosity was found to be $1.12^{+0.06}_{-0.07}\times 10^{42}$ erg
s$^{-1}$ for S1 and $7.7^{+0.4}_{-1.8}\times 10^{41}$ erg s$^{-1}$ for S2.
All these results are consistent with those obtained for previous outbursts in
which the thermal component dominated the spectrum during the outburst.

\section{Observational constraints put on the eccentric binary scenario}

Observations of the last two (2012 \& 2013) outbursts of HLX-1 provide
constraints on the eccentric-orbit donor model proposed by L11. First, the
possible 1--day delay between the rise in optical and that in X-rays (Webb et
al. 2014) can be explained in the framework of the mass-transfer model (MTM)
only if the orbit is very nearly parabolic. The standard explanation for the
long/short wavelength delay (see e.g. Pringle, Verbunt \& Wade 1996) is that
it is due to the density contrast propagating inwards, first through cool disk
regions. The observed delay precludes a viscous propagation because, in this
case, during one day the contrast would move just a couple of Schwarzschild
radii at best. Assuming a propagation at the speed of sound gives an upper
limit to the distance of $\sim 3\times 10^{11}$\,cm.  If this is the distance
where mass lost by the stellar companion is delivered to the disc, {\it i.e.}
roughly the periapsis distance, the implied eccentricity is so close to 1.0
that the orbit can be considered parabolic.  This issue has been anticipated
by L11 who wrote: ``{\it the actual response of a standard accretion disk to
  bursts of mass transfer may be too slow to explain the observations unless
  the orbit is close to parabolic and/or additional heating, presumably linked
  to the highly time dependent gravitational potential, is invoked.}" Such a
nearly parabolic, $e \rightarrow 1$, orbit can be unstable and disrupted
within several orbits, as we investigate with simulations in
Section~\ref{SPH}.

The second constraint arises from the recurrence time being variable (from 350
to 370 days between 2009 and 2012) and having increased significantly in 2013.
Even though the most recent outburst cycle was more than a month delayed
compared to the preceding four, all the outbursts are apparently of the same
type. These features can be explained in the framework of the eccentric-donor
MTM only if the orbital parameters can change significantly from orbit to
orbit.  Indeed, such a varying orbital period is the expected behavior for
certain binary systems. It is possible to either add or remove orbital energy
from the system depending on the oscillation phase of the secondary at
periapsis, and the resulting changes in orbital period can proceed in a
stochastic fashion (Mardling 1995a,b).  The criterion that the evolution is
stochastic is that the orbital period can change in one orbit by more than one
oscillation period of the secondary (Ivanov \& Papaloizou 2004, 2007).

\section{Smooth Particle Hydrodynamical simulations}
\label{SPH}
\subsection{Numerical set-up and parameter space}
\label{setup}

To investigate this possibility for HLX-1, we use the code {\it Starsmasher}
(Gaburov et al. 2010) to perform Newtonian smoothed particle hydrodynamical
simulations of the donor initially on a parabolic incoming trajectory towards
the BH.  The BH is modeled as a point particle, and the donor as a polytrope
with $N\approx 5\times 10^4$ particles.  We are aware that the use of less
  than $10^5$ particles corresponds to low resolution SPH simulations. However
  in this work, we want to follow a large number of dynamical timescales (from
  6000 to 70000) and to explore a large parameter space. This would not have
  been possible in large $N$ simulations. We do investigate the effects of
  particle resolution on our results by varying $N$ from $2.5\times 10^4$ to
  $2\times 10^5$ particles in one scenario. We find that the effects of using
  a higher particle resolution do not change the qualitative behavior of the
  system that we present in the following sections (see
  Section~\ref{resolution} for more details). 

We vary the structure of the donor via the polytropic index $n$, and we vary
the equation of state via the adiabatic index $\Gamma$.  Our polytrope models
are (a) $n=1.5$, $\Gamma=5/3$, (b) $n=2$, $\Gamma=1.5$, (c) $n=2$,
$\Gamma=5/3$, and (d) $n=3$, $\Gamma=5/3$.  By including these multiple donor
models in our study, we gain a better understanding of effects due to the
donor structure and equation of state.  Polytrope models of type (a) above are
appropriate for low mass main sequence stars, pre-main sequence stars, brown
dwarfs, low-mass stripped giant cores, and low-mass white dwarfs.  Polytrope
models of type (d) are reasonable representations of moderately massive
main-sequence stars like our Sun.  Realistic models of white dwarfs with
masses in the range from $0.9M_\odot$ to $1.3M_\odot$ have moments of inertia
that correspond to polytropes with $n$ between about 1.8 and 2.2 (Andronov \&
Yavorskij 1990); therefore polytrope types (b) and (c) provide a simple
approximation for the mass distribution in moderately massive white dwarfs or
stripped giant cores.  Given the scale-free nature of polytropes, each
simulation represents a family of cases, and we do not need to choose a
physical mass or radius of the donor until after a simulation when comparisons
with HLX-1 are made.  We survey parameter space by varying the mass ratio $q =
M/M_{\rm BH}$ from $10^{-3}$ to $10^{-5}$ and by varying the periapsis
separation $r_{\rm p}$ from $2.2~r_{\rm t}$ to $2.7~r_{\rm t}$, where the
tidal radius $r_{\rm t}=q^{-1/3}R$ and $M$ and $R$ are the initial mass and
radius of the donor, respectively. For fixed $r_{\rm p}/r_{\rm t}$, varying
$q$ has only a little effect beyond a rescaling of timescales, as expected for
$q<<1$.

\begin{figure}[ht!]
\includegraphics[scale=0.4]{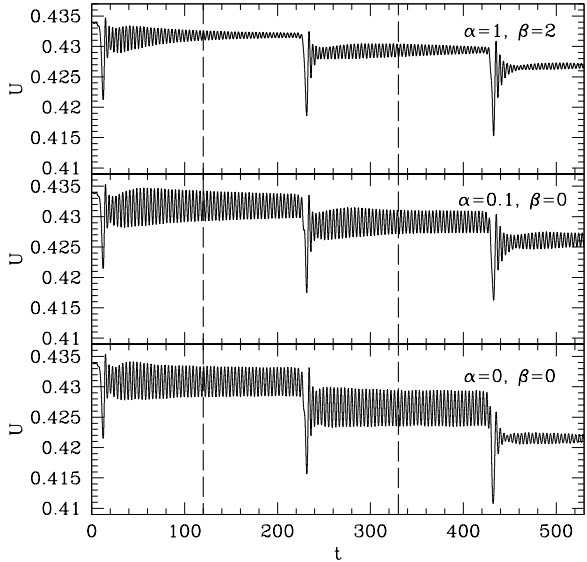}
\caption{Total internal energy $U$ versus time $t$ for three test simulations
  that differ only in their artificial viscosity (AV) parameters $\alpha$ and
  $\beta$.  Vertical dashed lines mark when the orbit is advanced via the
  Kepler two-body result.  The impulse at each periapsis passage (at times
  $t\approx 10$, 220 and 420) changes the oscillation state of the star.
  Local maxima in $U$ correspond to a maximally compressed star, while local
  minima correspond to maximal expansion.  In the top panel, stellar
  oscillations are quickly damped when the AV parameters $\alpha=1$ and
  $\beta=2$ are used.  Oscillations are much better preserved in the
  $\alpha=0.1$, $\beta=0$ and especially the $\alpha=0$, $\beta=0$
  simulations. The star is an $n=1.5$, $\Gamma=5/3$ polytrope, the inverse
  mass ratio $q^{-1}=M_{\rm BH}/M=3\times10^4$, and the initial periapsis
  separation is $r_{\rm p}=2.3\,r_{\rm t}$.  The unit of energy is $G M^2
  R^{-1}$, while the unit of time is $(GM)^{-1/2} R^{3/2} $, where $M$ and $R$
  are respectively the initial mass and radius of the star.  The production
  runs in this paper use $\alpha=0$ and $\beta=0$.
\label{AV}}
\end{figure}

For convenience and computational efficiency, we use the SPH method to model
only the close interactions during periapsis passages.  Once the donor has
retreated sufficiently far from the black hole to become stabilized (typically
about 100 dynamical timescales after periapsis), we employ the analytic Kepler
two-body result to advance the orbit to the same separation but now with the
donor infalling toward the BH (see Antonini et al. (2011) for more details on
the implementation of this method).  We allow for a random fluctuation in how
much time each periapsis passage is simulated, in order to randomize the
oscillation phase of the secondary when it returns to periapsis and therefore
to avoid the suppression of any stochastic behavior in the orbit.  During this
advancement of the orbit, we excise from the simulation any particles that
have been stripped from the star. In this way, we are guaranteed that any
changes in the orbital parameters that occur during subsequent periapsis
passages are due to mass transfer and tidal effects, and not due to
interactions between the donor and an under-resolved accretion disk.

If not handled carefully, artificial viscosity (AV) could spuriously damp
oscillations excited in our star and spuriously transport angular momentum
within it, affecting its size, structure, and rotational profile (Lombardi et
al.\ 1999).  We performed test calculations with different values of the
parameters $\alpha$ and $\beta$ used in the AV prescription described in the
appendix of Ponce et al.\ (2012).  Results of these calculations are shown in
Figure \ref{AV}, where we see that by decreasing the strength of the AV we
better preserve the oscillations induced by the periapsis passages. Note, for
example, the rather steady decrease in the amplitude of oscillations after the
first periapsis passage in the $\alpha=1$, $\beta=2$ calculation.  Given the
absence of shocks in our simulations, there is no real need for AV, and we
simply turn off AV completely by setting $\alpha=0$ and $\beta=0$ for the
production runs presented in this paper.  Doing so requires that smaller
timesteps be taken, but the payoff is that the oscillation modes and
rotational profile of the star are better modeled.

\subsection{Evolution of the orbital parameters}
\label{evolution}

\begin{figure*}[ht!]
\includegraphics[scale=0.85]{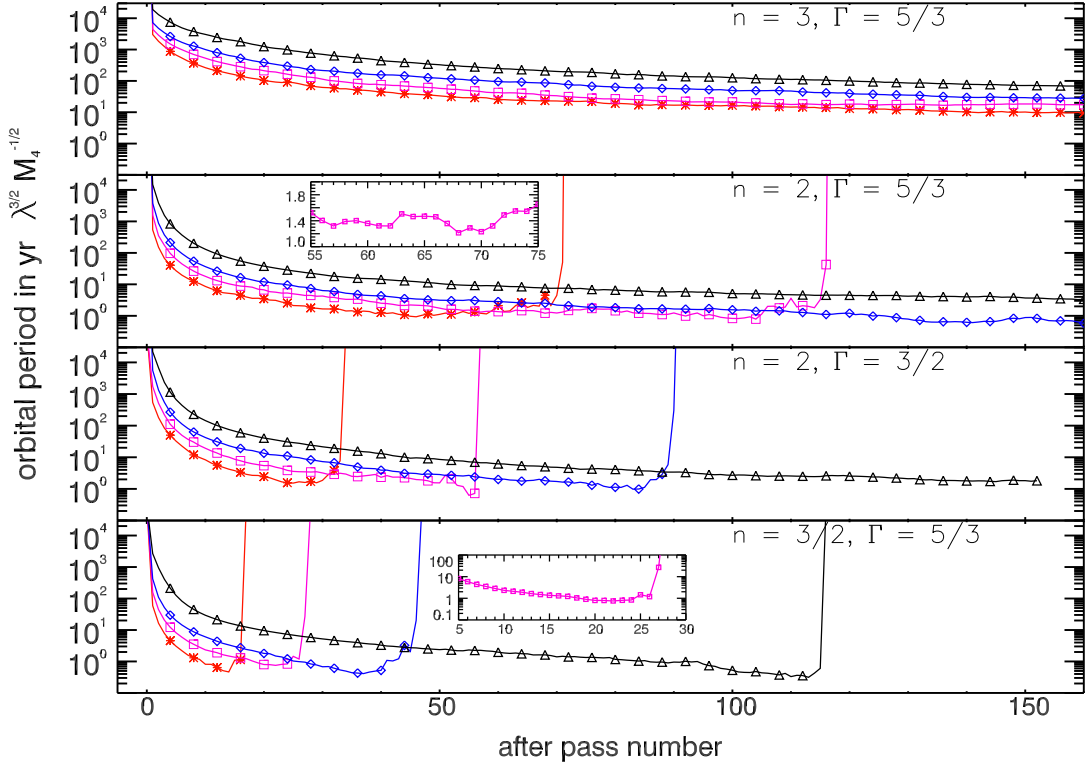}
\caption{Orbital periods after periapsis passages for simulations with various
  donor models and initial periapsis separations.  All donor models considered
  are polytropes: $n=1.5, \Gamma=5/3$; $n=2, \Gamma=1.5$; $n=2, \Gamma=5/3$;
  and $n=3, \Gamma=5/3$.  For each type of polytrope, the results for separate
  initial periapsis separations are shown, namely $r_{\rm p}/r_{\rm t}=2.3$,
  $2.4$, $2.5$, and $2.7$ (red stars, magenta squares, blue diamonds, and
  black triangles, respectively). Close-ups are shown for $r_{\rm p}/r_{\rm
    t}=2.4$ for two polytropes: $n=1.5, \Gamma=5/3$ and $n=2 \Gamma=5/3$.  In
  all cases, the inverse mass ratio $q^{-1}=3\times 10^4$.  Curves terminate
  on their right edges either because the orbit becomes hyperbolic or the
  simulation was terminated.  The units of the orbital period are
  yr\,$\lambda^{3/2}M_4^{-1/2}$, where $\lambda\equiv R/(0.01R_\odot)$ and
  $M_4\equiv M_{\rm BH}/(10^4 M_\odot)$.  Note how in the several cases, the
  orbital period maintains a roughly constant value of order 1 year, if
  $\lambda^{3/2}M_4^{-1/2}\sim 1$, before increasing.
\label{periods}}
\end{figure*}

We find that several of our simulations with $r_{\rm p}/r_{\rm t}$ in the 2.2
to 2.7 range exhibit behavior in the pattern of orbital periods that is
qualitatively consistent with that observed in HLX-1.  This behavior is shown
in Table \ref{sph_table} and Figure \ref{periods} for cases with inverse mass
ratio $q^{-1}=3\times 10^4$.  For example, notice that during periapsis
passages 17 through 24 of the $n=1.5$, $\Gamma=5/3$, $r_{\rm p}=2.4\,r_{\rm
  t}$ case (see the close-up in the bottom panel of Fig.~\ref{periods}),
the orbital period passes through a shallow minimum, maintaining an
approximate equal value for several orbits.  After periapsis passage 25, the
orbital period increases by more than 10\% (much more in this case), in
qualitative accord with the recent delayed outburst of HLX-1 (see Table
\ref{sph_table}).  Here, mass transfer is starting to become substantial, with
slightly more mass in the tail stretching toward the black hole than in the
tail extending outward.  This asymmetry causes the orbit to increase in
semi-major axis and eccentricity until ultimately becoming unbound, an effect
previously observed in similar simulations (Faber et al.\ 2005, Manukian et
al.\ 2013).

An inspection of Figure~\ref{periods} reveals an even more shallow minimum in
the $n=2$, $\Gamma=5/3$, $r_{\rm p}=2.4\,r_{\rm t}$ case (see the close-up in
the second panel from the top in Fig.~\ref{periods}), with the curve actually
containing two extended local minima due to the chaotic nature of the energy
transfer during periapsis passages (Mardling et al.\ 1995a,b).  The emergence
from these minima occur after passages 63 and 72, with a more than 10\%
increase in orbital period in each case.  Interestingly, after the subsequent
passages 64 and 73, the period decreases in one case and increases in the
other.  As in the corresponding $n=1.5$ case, the system emerges from the
global minimum once mass transfer becomes substantial enough to give the orbit
a boost to a larger semi-major axis and eccentricity.  Within several orbits,
the donor becomes unbound from the BH.  If a similar mechanism is indeed
responsible for the orbital period variation in HLX-1, then (by definition of
``stochastic") it is not possible to predict whether the next outburst cycle
will be longer or shorter than the last.  However, if HLX-1 is emerging from a
shallow minimum in orbital periods, the general pattern to expect would be for
the duration of the outburst cycle to tend to increase until the donor is
ejected and the mass transfer episodes stop altogether.

Periapsis separations outside of the range explored in Figure~\ref{periods}
are interesting to consider.  Situations with $r_{\rm p}\gtrsim 3\,r_{\rm t}$
correspond to the region of parameter space explored in the classic tidal
capture process (see, e.g., Novikov et al.\ 1992, Hopman et al.\ 2004,
Baumgardt et al.\ 2006).  However, these situations would not provide the mass
transfer events necessary to explain the quasi-periodic HLX-1 outbursts.  Our
simulations with $r_{\rm p} \le 2.1\,r_{\rm t}$ are not consistent with the
observed properties of HLX-1. Although mass stripping and tidal effects during
the initial periapsis passage cause the orbit to become bound (while still
highly eccentric), subsequent passages disrupt the donor enough to change the
orbital parameters too substantially from one orbit to the next.  In addition,
the stochastic fluctuations are too pronounced, when compared to those seen in
the cases with larger $r_{\rm p}$.

\subsection{Constraints on the donor type}
\label{donor}

To understand how our results scale with the mass ratio $q$, the orbital
parameters after one periaspsis passage can be determined quasi-analytically
using the method of Press \& Teukolsky (1977).  For our situations, the first
periapsis passage excites primarily the $l=2$ oscillation mode of the star, and
the energy transferred out of the orbit by the tidal interaction is given by
\begin{eqnarray*}
\Delta E = \frac{G M_{\rm BH}^2}{R}\left(\frac{R}{r_{\rm p}}\right)^6 T_2(\eta),
\end{eqnarray*}
where the dimensionless function $T_2(\eta)$ is evaluated in Lee \& Ostriker
(1986) for three of the polytrope types considered in this paper and
$\eta=(r_{\rm p}/r_{\rm t})^{3/2}$.  Because our initial orbit is parabolic,
the semimajor axis after one passage is simply $a=G M_{\rm BH}M/(2\Delta
E)$. From Kepler's third law, the associated orbital period after the first
periapsis passage is
\begin{equation}
P=1.12\times 10^{-9} \, \eta^6 \left(\frac{\lambda}{T_2(\eta)
  q}\right)^{3/2}M_4^{-1/2}\,\hbox{yr}, \label{PT}
\end{equation}
where $\lambda\equiv R/(0.01R_\odot)$ is a dimensionless measure of the donor
radius and $M_4\equiv M_{\rm BH}/(10^4M_\odot)$ is a measure of the black hole
mass

In our simulations at a given $r_{\rm p}/r_{\rm t}$ and fixed $q<<1$, equation
(\ref{PT}) shows that the orbital period scales like $\left(R^3/(GM_{\rm
  BH})\right)^{1/2}$.  As $q$ varies, we confirmed through simulations that
the orbital periods through multiple passages scale like $q^{-3/2}$ for a
given initial $r_{\rm p}/r_{\rm t}$ (smaller $q$ means larger orbits and
longer orbital periods).  Our results indicate that the minimum orbital period
satisfies

$$P_{\rm min}\approx p\left(\frac{\lambda}{3\times 10^4
  q}\right)^{3/2}M_4^{-1/2}\,\hbox{yr}.$$ Here $p$ is simply a numerical
coefficient whose value depends on the type of donor and can be read off from
the vertical axis of Figure~\ref{periods} by looking for the minimum in the
curve corresponding to the donor star under consideration.  For example, for
an $n=1.5$ donor, $p\approx 0.3$ to 0.8. For our $n=2$ polytropes, $p \approx
0.6$ to 1.5.  In all of our simulations, $p$ is never less than $\sim 0.3$
(see Fig.~\ref{periods}).  Although prohibitive computational costs prevent
us from following all curves in Figure~\ref{periods} to their minima, the
trends in our results are consistent with there being a minimum for any
combination of donor star and initial $r_{\rm p}$, with the value of $p$ being
more sensitive to the polytropic index $n$ than to the adiabatic index
$\Gamma$ or periapsis separation $r_{\rm p}$.  Additional test calculations
done with artificial viscosity support this inference.

If the delayed outburst of HLX-1 is due to its emergence from the minimum of a
curve like that shown in Figure \ref{periods}, then $P_{\rm min}\approx
1$\,yr, or, after some algebraic manipulation,

\begin{equation}
1 \approx p \left(\frac{\lambda}{3\mu}\right)^{3/2}M_4,\label{mr}
\end{equation}
where $\mu\equiv M/M_\odot$ is a dimensionless measure of the initial donor mass.

Spectral modeling and Eddington scaling from X-ray/radio data suggest that
$M_{\rm BH}\gtrsim10^4 M_\odot$ (or equivalently $M_4\gtrsim 1$). This then
provides stringent requirements on the possible mass $M$ and radius $R$ of the
donor.  In particular, the expression on the right hand side of equation
(\ref{mr}) is much greater than 1 for any MS star, pre-MS star, brown dwarf,
or planet, ruling out the possibility of such donors in this model.  The right
hand side can, however, be of order unity for white dwarfs or stripped giant
cores. White dwarfs obey an approximate mass-radius relation $MR^3\approx
6\times 10^{-7}\,M_\odot\,R_\odot^3$, or equivalently $\mu\lambda^3\approx
0.6$, at moderate masses.  Using this mass-radius relation for white dwarfs in
equation (\ref{mr}) yields
\begin{equation}
M\approx 0.4 M_\odot p^{1/2} M_4^{1/2}. \label{wd_result}
\end{equation}

A BH of mass $\sim 2\times 10^4 \,M_\odot$, for example, could tidally strip
matter from an $n=1.5, \Gamma=5/3$ WD of mass $M\sim 0.5\,M_\odot$ and yield
mass transfer episodes at a rate of one per year.  The $n=2$ polytropes have
larger $p$ and therefore require a larger WD mass for the same $P_{\rm min}$:
for example, $p\approx1.5$ implies $M\approx 0.7M_\odot$ for $M_4\approx2$.  A
more massive black hole would require a more massive white dwarf to achieve
the same $P_{\rm min}\approx 1$yr. 

 Tidal disruption of white dwarfs by IMBHs will lead to fast rising, highly
 energetic events, that should be observable more often than tidal disruption
 of main-sequence stars (Shcherbakov et al.\ 2012, MacLeod et al. 2014).
 Simulations by Rosswog et al.\ (2009) of WDs and moderately massive BHs show
 that it is possible to trigger a nuclear explosion that would destroy the WD,
 as may have been the case in GRB060218 (Shcherbakov et al.\ 2013).  However,
 in the simulation of Rosswog et al.\ (2009) that most closely resembles those
 presented here (see run 12 in their Table 1), the WD is not compressed enough
 to trigger explosive nuclear burning.

\subsection{Discussion on the mass accretion rate}

Consider a system with an $M=0.5\,M_\odot$, $R=7\times 10^8$\,cm, $n=1.5,
\Gamma=5/3$ WD donor and an $M_{\rm BH}=2\times 10^4\, M_\odot$ BH interacting
with periapsis separation $r_{\rm p}=2.4\,r_{\rm t}=6\times10^{10}$ cm
(corresponding here to $\sim 10~R_g$ with $R_g$, the gravitational radius).
In this picture, the delayed outburst of 2013 would be equated with the
periapsis passage represented by row 25 in Table \ref{sph_table}, when the
orbital period actually increases by substantially more than 10\%.  In that
passage, about 2\%, or $\sim 10^{-2}\, M_\odot$, of the mass is stripped from
the WD donor.  A little more than half of this stripped material falls toward
the black hole: $M_{\rm fallback}=0.0106M=0.0053M_\odot$. In the simulations
we ran, we are unable to follow the accretion of the bound matter onto the
BH. The way this matter is accreted following each periapsis passage will
depend on several parameters such as the BH spin, the orientation of the
BH spin with respect to the system orbital plane (see e.g. Haas et
al. 2012). If matter is delivered sufficiently close to the BH, general
relativity effects can also become important to consider (see e.g. Dai et
al. 2013).
 
\begin{figure}[ht!]
\includegraphics[angle=0,scale=0.4]{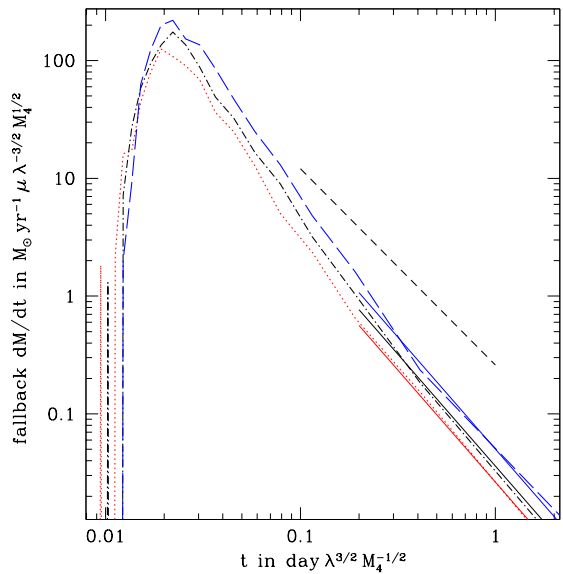}
\caption{Mass fallback rate versus time after the 24th (red dotted curve),
  25th (black dot-dashed curve), and 26th (blue long dashed curve) periapsis
  passages in the $n=1.5$, $\Gamma=5/3$, $r_{\rm p}=2.4~r_{\rm t}$ simulation.
  The thin solid lines at late times give the result of equation (\ref{mdot}),
  with its assumed $t^{-1.9}$ dependency, for each of the passages.  For
  comparison, the short dashed curve shows a $t^{-5/3}$ dependency.  The
  spikes near $t=0.01$\,day$\lambda^{3/2}M_4^{-1/2}$ are simply fluctuations due
  to single SPH particles.
\label{fallback}}
\end{figure}

The stripped matter has first to lose its large angular momentum before being
able to be accreted by the BH. If the timescale over which most of the bound
matter loses its angular momentum is sufficiently short, then a small thick
accretion disk may form around the BH. Assuming that the dynamics of the
stripped matter is the same as that predicted in tidal disruption events
(TDEs), we estimate that the fallback time is less than $\sim 1$ hour using
the formalism developed in Donato et al. (2014) \& Rosswog et al. (2009). So,
in this case a thick disk is expected to form rapidly (i.e. a few times the
fallback time). Then, matter in the disk will diffuse inwards over the viscous
timescale in order to be accreted by the BH. In order to have a rough estimate
of the accretion rate, we use the freezing model of MacLeod et al.\ (2012) and
Guillochon \& Ramirez-Ruiz (2013) as
$$\frac{dM}{dt}=\frac{dM}{dE}\frac{dE}{dt}=\frac{1}{3}\left(2\pi G M_{\rm
  BH}\right)^{2/3}\frac{dM}{dE}t^{-5/3},$$ where Kepler's third law has been
used to relate the specific orbital binding energy $E=GM_{\rm BH}/(2a)$ to the
time $t$ for material to return to periapsis.\footnotemark\footnotetext{Note:
  we find the same numerical coefficient in our equation for $dM/dt$ as in
  MacLeod et al.\ (2012), which is different than that given in Guillochon \&
  Ramirez-Ruiz (2013).}

Figure \ref{fallback} shows the fallback rate $dM/dt$ for the 24th through
26th passages. The peak fallback time is equal to $t_{\rm peak}\sim 0.02 {~\rm
  day} \lambda^{3/2} M_4^{-1/2}$. The peak fallback time is only weakly
dependent on the donor type and on, for our close encounters, the orbital
parameters.  The peak fallback rate of the 25th passage, corresponding to the
peak of the dot-dashed curve in Figure~\ref{fallback}, is $(dM/dt)_{\rm peak}
=175 \mu\lambda^{-3/2}M_4^{1/2}$ (incidentally, the coefficient of 175 is one
of the entries in Table~\ref{sph_table}).  For a WD orbiting in HLX-1, this
corresponds to $(dM/dt)_{\rm peak} \approx 10^2 M_\odot\,{\rm yr}^{-1}$, which
is orders of magnitude more than the Eddington limit, $\dot M_{\rm Edd}=2.1
\times 10^{-5} \epsilon^{-1} M_4 M_\odot {\rm yr}^{-1}$ with $\epsilon$, the
radiative efficiency ($\dot{M}_{\rm Edd} \sim 7 \times 10^{-4}~M_\odot$
yr$^{-1}$ for a $2\times 10^4\,M_\odot$ BH accreting with an efficiency of
$\epsilon = \frac{1}{16}$).  Our simulation results for the fallback rate at
times $t$ well after $t_{\rm peak}$ can be approximated as
\begin{equation}
\frac{dM}{dt} \approx 3.4 M_{\rm fallback} \left(\frac{t}{\rm day\lambda^{3/2}M_4^{-1/2}}\right)^{-1.9} \lambda^{-3/2} M_4^{1/2} {\rm yr}^{-1}, \label{mdot}
\end{equation}
where $M_{\rm fallback}$ is the total mass during that periapsis passage that
falls back toward the BH.

\begin{figure*}[ht!]
\includegraphics[scale=0.8]{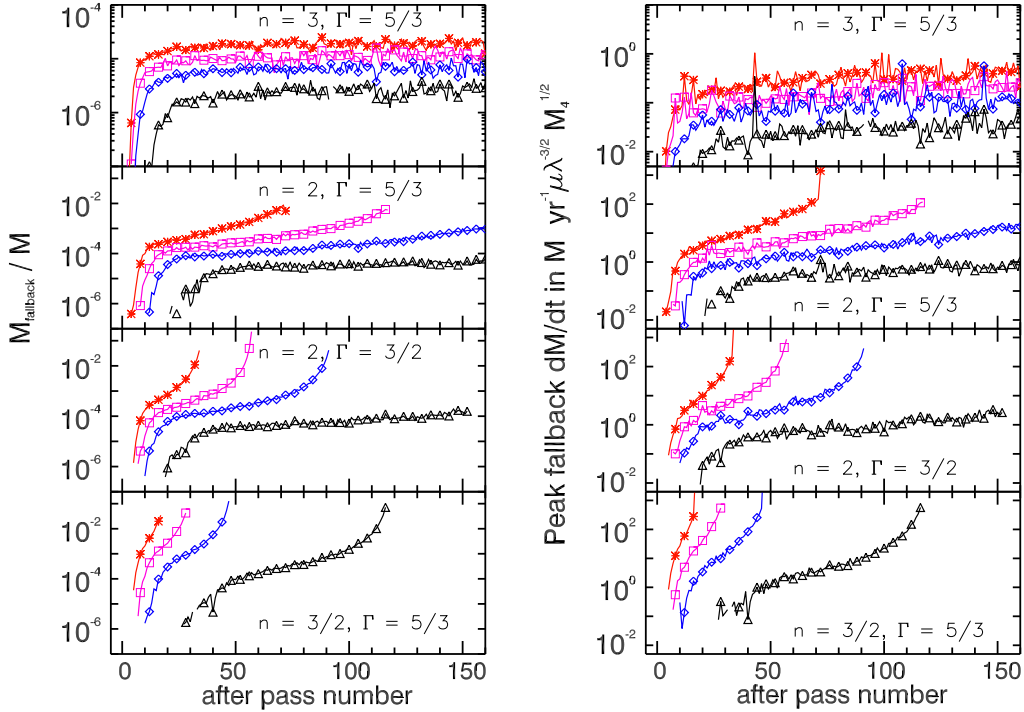}
\caption{{\it Left} -- Ratio of the fallback mass over the donor mass; {\it
    Right} -- the peak fallback rate $(dM/dt)_{\rm peak}$.  Data point types
  are as in Fig.~\ref{periods}.
\label{mass}}
\end{figure*}

From equation (\ref{mdot}), we calculate the time $t_{\rm Edd}$ after a
periapsis passage at which the accretion drops below the Eddington accretion
rate $\dot M_{\rm Edd}$:
\begin{equation}
t_{\rm Edd}\approx  550\,{\rm day} \left(\epsilon\mu \frac{M_{\rm fallback}}{M}\right)^{0.53} \lambda^{0.71} M_4^{-0.76}. \label{tEdd}
\end{equation}
Equation (\ref{tEdd}) is used to provide the last column of information in
Table \ref{sph_table}.  So, the fallback rate remains super-Eddington for only
a few days.  For other simulated scenarios with less fallback mass, the time
until the accretion becomes sub-Eddington can be an order of magnitude
smaller.

In Fig.~\ref{mass}, we show the fallback mass $M_{\rm fallback}$ and the peak
fallback $(dM/dt)_{\rm peak}$ after each periapsis passage for all of the
simulations presented in Fig.~\ref{periods}. In all cases shown in
Fig.~\ref{mass} and for the type of donor we consider here (a white dwarf or a
stripped giant core), a brief super-Eddington accretion episode is expected to
take place following the formation of a thick disk shortly after each
periapsis passage. Depending on the peak accretion rate, these episodes could
occur within a day making the observation of such events rather challenging.
The presence of powerful outflows, often invoked in the super-Eddington
accretion regime (e.g. Ohsuga et al.\ 2005, Poutanen et al. 2007, Ohsuga \&
Mineshige 2011, Ohsuga \& Mineshige 2013; see also De Colle et al.\ 2012),
might make this phase even shorter by depleting matter from the disk. Such
outflows coupled with the unbound stripped matter might even be able to power
a nebula around the system. The $t_{\rm Edd}$-values quoted in
Table~\ref{sph_table} may be considered as upper limits.  Super-Eddington
accretion is predicted to give rise to powerful electromagnetic
emission. However, this emission may be highly anisotropic (e.g. King et
al. 2001, Kawashima et al. 2012, Ohsuga \& Mineshige 2011, 2013). In some
cases, this may even prevent observing direct emission from the inner parts of
the disk if the observer line of sight is sufficiently far from the disk
axis. The observed emission flux may be then significantly decreased, making
the observation of such extreme events even more challenging.

Once the accretion rate is close to or below the critical value, most of the
outflows might switch off (except maybe within the innermost parts of the
disk), making possible the direct observation of the disk emission. This would
lead to a rapid increase in the source luminosity in X-rays and optical. In
such a picture, the super-Eddington accretion episodes would appear as
precursors to the main outbursts that would correspond to matter accreted at a
rate close to or below the Eddington limit. We did not find any strong
evidence in the observational data collected so far for the presence of a
precursor preceding the main outburst. However, this can be easily accounted
for in our observing strategy to monitor the change in luminosity in HLX-1,
since the shortest timescale used when the source was in the low/hard state
was to collect a 1\,ks snapshot every 2 days.

 In Fig.~\ref{mass}, the general pattern is that the peak fallback rate tends
 to increase over several periapsis passages, even if from passage to passage
 the value of the accretion rate could vary in a stochastic manner due to the
 effects of stochastic fluctuations on the star (see
 Section~\ref{stochastic_evolution}). Even if this seems to be difficult to
 reconciliate with our observations, how this translates in term of outburst
 duration deserves a proper investigation that is beyond the scope of the
 present paper.  Another important point to consider in future works is that
 in the case of HLX-1 we know from the X-ray observations that there is still
 some material accreted when the source is in the low/hard state. How this
 material and the matter transferred from the star at periapsis interact to
 give rise to an accretion disk around the BH also deserves detailed scrutiny.

\subsection{Discussion on the lifetime of such a highly eccentric system}
\label{life}

\begin{figure}[ht!]
\hspace{-1cm}\includegraphics[scale=0.5]{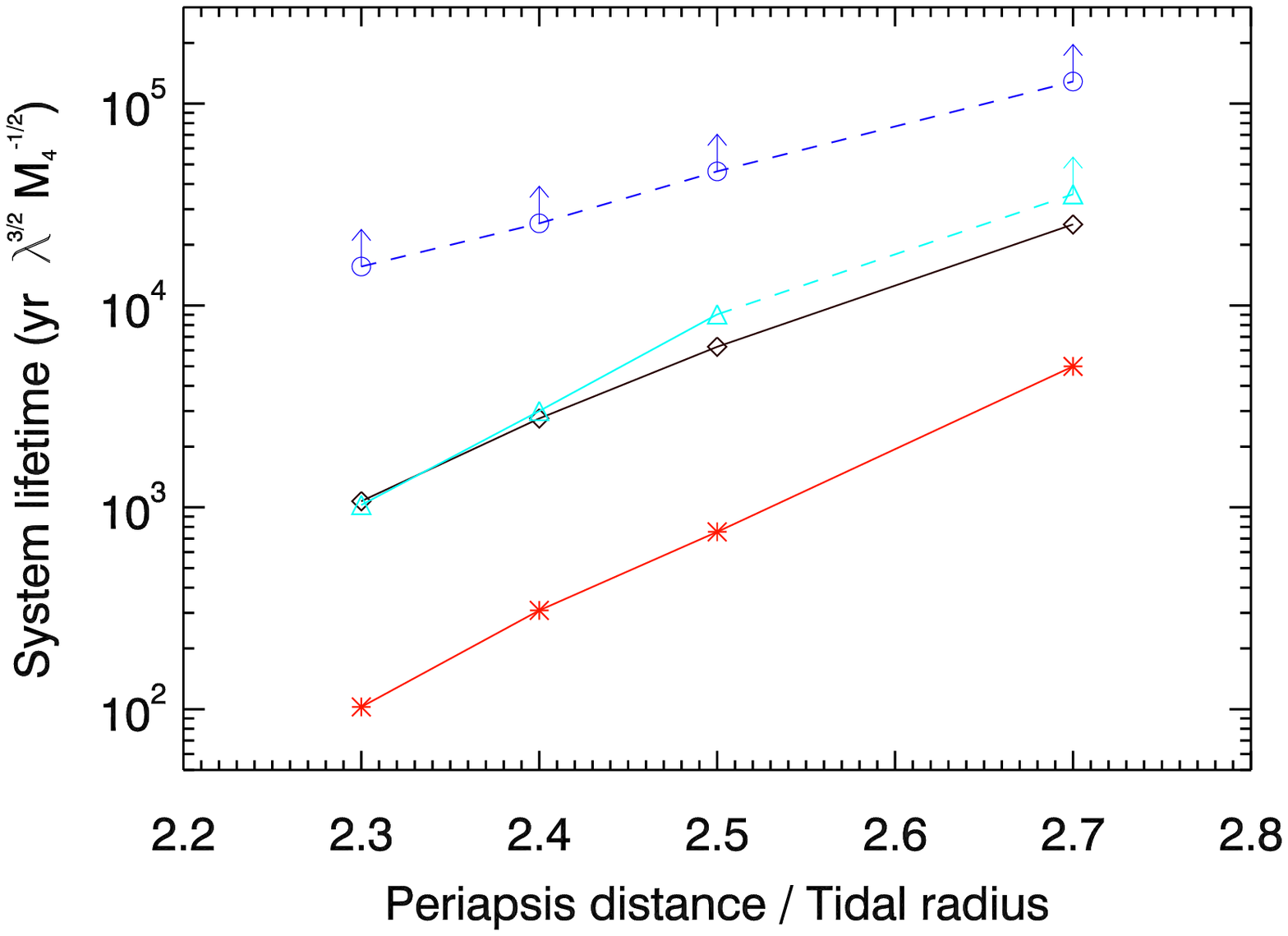}
\caption{Lifetime of the system for the different cases shown in
  Figure~\ref{periods}: $n=1.5, \Gamma=5/3$ (red stars), $n=2, \Gamma=5/3$
  (black squares), $n=2, \Gamma=1.5$ (cyan triangles) and $n=3, \Gamma=5/3$
  (blue circles). The data points for which the simulation was stopped before
  the star becomes unbound to the BH because of too prohibitive computation
  time are marked with vertical arrows. Therefore, they correspond to lower
  limits on the lifetime of the system.
\label{fig_life}}
\end{figure}

Figure~\ref{fig_life} shows the lifetime of the system ($\Delta t_{\rm sys}$)
for the different cases shown in Figure~\ref{periods}. The lifetime of the
system strongly depends on how efficient is the tidal stripping of the donor
star. Thus, the distance at periapsis and the composition of the donor star
(i.e. the equation of state of the polytrope considered) have a significant
impact on $\Delta t_{\rm sys}$. Smaller $r_p /r_t$-values imply that more
matter is transferred towards the BH at a given periapsis passage (see
Fig.~\ref{mass}). This results in a quicker way to make the star unbound to
the BH (see Section~\ref{stochastic_evolution}). In all cases considered in
this paper, the $\Delta t_{\rm sys}$-values appear to be rather short. For the
case of the WD donor around a $2\times 10^4~M_\odot$ BH considered in the
above sections (i.e. $r_p /r_t = 2.4$, $\lambda = 1$, $n=1.5$ \& $\Gamma =
5/3$), $\Delta t_{\rm sys} \sim 220$\,years; which makes the detection of such
a system very challenging. Note the lifetime of the system will even be
smaller when considering higher BH mass since $\Delta t_{\rm sys} \propto
M_{\rm BH}^{-1/2}$.

Obviously, the lifetime of the system is not related to the time interval over
which the source emission will be detectable. The latter time depends on when
mass transfer starts and how much matter remains in the disk before the star
is ejected. As seen in Fig.~\ref{mass} (black triangles i.e. $r_p /r_t =
2.7$), mass transfer episodes do not necessarily start once the star becomes
bound to the BH. During the first periapsis passages, the star radius is too
small for mass transfer to take place. Once the star radius increases enough
due to the star oscillations induced by tidal forces over successive periapsis
passages, mass transfer episodes will start. When mass transfer episodes stop
because the star has been ejected, the source should stay in the low/hard
state until the source will run out of fuel. Once this happens, then the
source will switch into quiescence. According to this scenario, when the donor
star will be ejected, HLX-1 should look like a moderately bright ULX in the
outskirts of the ESO 243-49 S0 galaxy with an X-ray luminosity of $\sim
10^{40}$ erg s$^{-1}$ displaying spectral properties consistent with some
known ULXs.

\subsection{Discussion on the stochastic orbital evolution}
\label{stochastic_evolution}

\begin{figure}[ht!]
\hspace{-1.cm}\includegraphics[scale=0.6]{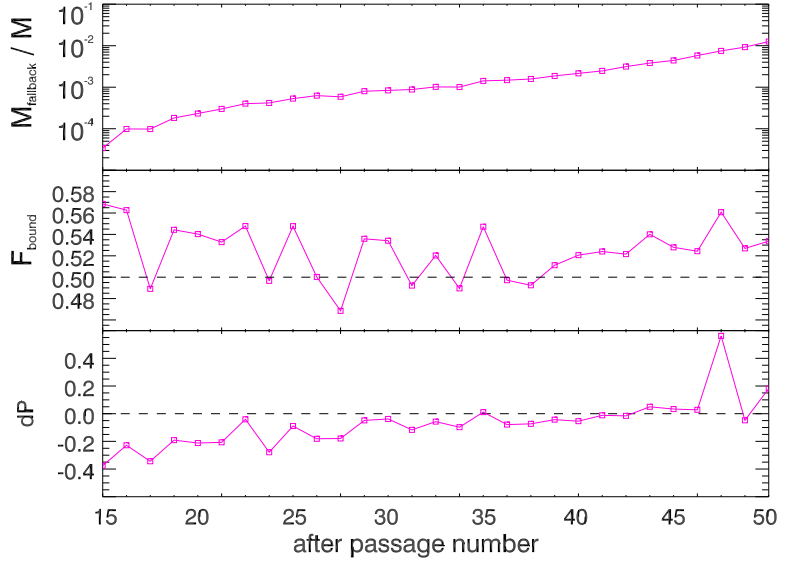}
\caption{Effects of stochastic fluctuations on the orbital evolution. Top --
  Ratio of the fallback mass to the initial mass of the star; Middle --
  Fraction of the stripped matter bound to the BH; Bottom -- change in the
  orbital period from one orbit to the next. The three curves are shown for
  the case: $r_p / r_t = 2.5$, $n=1.5$ \& $\Gamma=5/3$ with artificial
  viscosity turned off. $dP$ is given in units of yr
  $\lambda^{3/2}~M_4^{-1/2}$.
\label{fstochastic}}
\end{figure}

Our simulations suggest the pattern in orbital period and outbursts in HLX-1
could be explained by the combination of mass transfer and stochastic orbital
evolution. This is illustrated in Figure~\ref{fstochastic} comparing the
evolution over time of the changes in orbital period from one orbit to the
next ($dP$) and the fraction of mass falling onto the BH ($F_{\rm
  bound}$). The curves are shown for the case: $r_p / r_t = 2.5$, $n=1.5$ \&
$\Gamma=5/3$ with artificial viscosity turned off. The passages with $F_{\rm
  bound} > 50\%$ of the mass falling onto the black hole result in a smaller
decrease in orbital period than for those passages with $F_{\rm bound}\le
50\%$. This could be understood as mass loss falling onto the BH ``{\it
  pushes}'' the orbit to larger semimajor axis, while mass loss towards
infinity ``{\it pushes}'' the orbit to smaller semimajor axis. Similar
behavior is seen for the other cases considered in this paper. It appears that
stochastic behavior in mass loss is tied to stochastic behavior in the orbital
period.

The appearance of stochastic behavior in mass loss are due to the development
of stochastic fluctuations inside the star. For most cases considered here,
the criterion for stochastic fluctuations is satisfied from the first orbits;
which means that the phase of the oscillation is essentially randomized at the
next periapsis passage. Most orbital energy gets dumped into the $l=2$ mode of
oscillation.  This does not imply that stochastic fluctuations will
necessarily have large effects on the behavior of the orbital period at early
times. Once the star is sufficiently perturbed after several periapsis
passages, the randomness of the oscillation phase at periapsis results in a
randomness in which sides the star will be bulged out at periapsis, and
therefore the amount of mass stripping will depend on which way the star
happens to be bulging at periapsis. When the amount of mass stripping becomes
substantial at later times, this could lead to large positive changes in the
orbital period, making the system less bound and ultimately the star will be
ejected.

If the energy is dumped into many modes of oscillation, then the stochastic
effects on the orbit are diminished (e.g. Kochanek 1992).  Each mode of
oscillation has its own oscillation period, and whether there is energy put
into the orbit or taken out of the orbit, from one particular mode, depends on
the relative phase of the oscillation and the orbit at periapsis.  If there
are many modes of oscillation that have been excited, then some of the modes
will remove energy from the orbit, while other modes will add energy to the
orbit, and these effects tend to cancel so that the stochastic behavior is
less evident.  If there are only a small number of modes that are
substantially excited, then the stochastic behavior is more evident because
when adding up the individual effects there are too few of them always to get
an overall cancellation.

Below, we apply the equations of Ivanov \& Papaloizou (2007, hereafter IP) to
understand better under what conditions the stochastic process can give the
necessary orbital period fluctuations.\footnote{It is worth noting that the
  theory of linear tides presented in IP is not intended to be applied once
  mass transfer and mass loss occur.  Still, the criterion for stochastic
  behavior can be expressed as the number of oscillations that the donor
  undergoes between periapsis passages changing by more than 1 from one orbit
  to the next.  This condition can equivalently be expressed as the phase
  change $\delta \Phi > 2\pi$, so that phase coherence is lost and the donor
  is in an essentially random phase with each successive orbit.  Mass transfer
  and mass loss alter not only the orbital period $P_{\rm orb}$ but also the
  fundamental oscillation frequency $\omega_f$ of the donor, so that $\delta
  \Phi= \omega_f \delta P_{\rm orb} + P_{\rm orb} \delta \omega_f$.  While
  $\delta \omega_f=0$ in the scenario treated by IP, here the non-zero $\delta
  \omega_f$ acts to make the region of parameter space within which stochastic
  behavior operates even larger than what the equations of IP imply.  } We
calculate the minimum separation at which the stochastic instability would
operate from equation (13) of IP:
\begin{eqnarray*} 
a_{\rm st} \approx \lambda\mu^{-3/5}M_4^{3/5}(\Delta \tilde{E})^{-2/5} 10^{11}{\rm cm},
\end{eqnarray*}
where the fractional change in orbital binding energy $\Delta \tilde{E}\equiv
\Delta E/E$ can be calculated from equation (7) in IP and the specific orbital
binding energy $E=GM_{\rm BH}/(2a)$.  Here we neglect the contribution of
gravity waves to $\Delta \tilde{E}$, which have a secondary effect on the
orbital evolution, at least for slowly rotating white dwarfs (see the
discussion in IP at the end of their \S 4).

For instance, consider a donor star orbiting a BH with a 1 yr orbital period,
as in our model for HLX-1.  By Kepler's third law, the semimajor axis $a=
3.2\times10^{14}$cm$\,M_4^{1/3}$.  For a WD secondary that satisfies the mass
radius relation $\mu \lambda^3\approx 0.6$ and that has a mass given by
equation (\ref{wd_result}) with $p\sim 1$, the condition $a>a_{\rm st}$
requires roughly $\Delta \tilde{E}\gtrsim 10^{-8}M_4^{-1/2}$.  We find for our
WD that $\Delta \tilde{E}\approx 20/(\phi \Psi)$, where $\phi$ is a function
of $\eta\equiv (r_{\rm p}/r_{\rm t})^{3/2}$ given by equation (8) of IP and
$\Psi$ is a rotation parameter given by their equation (12).  For $M_4\sim 2$
and a non-rotating star ($\Psi=1$), the condition $a > a_{\rm st}$ requires
$\eta\equiv (r_{\rm p}/r_{\rm t})^{3/2}\lesssim 9$ or $r_{\rm p}/r_{\rm
  t}\lesssim 4.3$.  For $M_4\sim 2$ and a star rotating near breakup, $\Delta
\tilde{E}\gtrsim 10^{-8} M_4^{-1/2}$ requires $\eta \lesssim 7$ or
equivalently $r_{\rm p}/r_{\rm t} \lesssim 3.7$.  We therefore should expect
stochastic behavior in all of our scenarios we are considering.

Should the stochastic fluctuations be large enough to explain a $\sim$10\%
change in orbital period?  A 10\% change in period requires a $\sim$7\% change
in semimajor axis or in orbital energy: $0.07\approx \Delta \tilde E$.

So, for a slowly rotating WD, a $\sim$10\% change in period requires $\phi
\sim 300$, which corresponds to $\eta\sim 3.5$ or $r_{\rm p}/r_{\rm t}\sim
2.3$.  We note that this rough estimate neglects orbital energy changes
associated with mass loss, so we expect 10\% fluctuations in orbital period
also to be obtainable in scenarios that had initial $r_{\rm p}/r_{\rm t}$
larger than 2.3.  As found in our simulations, the stochastic fluctuations in
orbits with initial $r_{\rm p}/r_{\rm t}$ in the 2.3 to 2.7 range are the
right order of magnitude to explain what is being observed in HLX-1.  Although
computational costs prohibit us from following situations with initial $r_{\rm
  p}/r_{\rm t}>2.7$ all the way down to the minimum in orbital period,
such scenarios may also provide large enough stochastic fluctuations to
explain the behavior of HLX-1.

\subsection{Effects of viscosity on the orbital period}
\begin{figure}[ht!]
\hspace{-2cm}\includegraphics[scale=0.6]{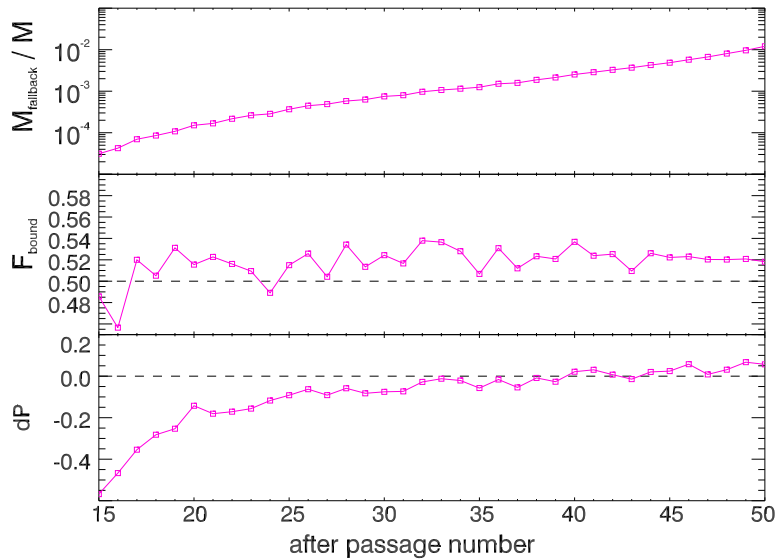}
\caption{Same legend as in Fig.~\ref{fstochastic}, but with artificial
  viscosity turned on ($\alpha = 1$ \& $\beta = 2$).
\label{stochasticAV}}
\end{figure}

\begin{figure}[ht!]
\hspace{0cm}\includegraphics[scale=0.5]{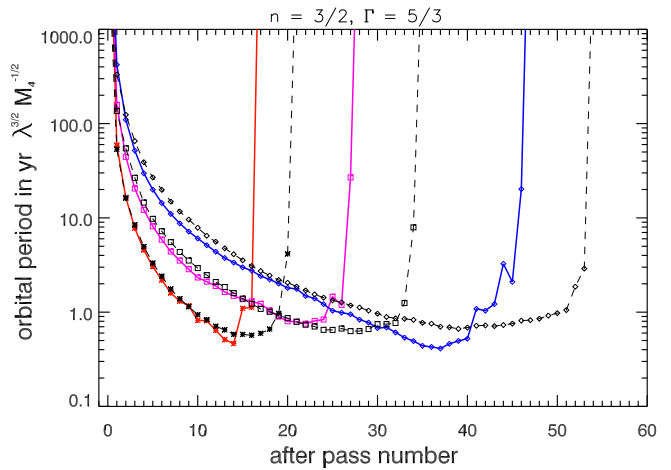}
\caption{Effects of viscosity on the evolution of the orbital period for the
  case: $n = 1.5$ \& $\Gamma = 5/3$ and different $r_p / r_t$ values: 2.3 (red
  stars), 2.4 (magenta squares) and 2.5 (blue diamonds). Thicks lines
  correspond to simulations run with AV turned off and dashed lines correspond
  to simulations run with AV turned on.
\label{periodAV}}
\end{figure}

We ran all our simulations with AV turned off. As shown in Fig.~\ref{AV}, the
effects of viscosity are to damp the tidal oscillations of the star so that the
effects of stochastic fluctuations are mostly suppressed.  This results in a
smoother behavior in mass loss and hence smoother changes in the orbital
period as seen in Figure~\ref{stochasticAV}. Due to the almost complete
suppression of stochastic fluctuations, the star will stay bound to the BH on
a longer timescale than in the case without AV (see
Fig.~\ref{periodAV}). However, the star is expected to be ultimately ejected
with and without AV. We note that at later times when the star is heavily
perturbed stochastic behavior in the orbital period could be observed in the
simulations we ran with AV turned on.

Do the effects of viscosity play an important role in damping the star
oscillations in the case where the donor is a white dwarf?  

Kochanek (1992) stressed that viscosity can damp the oscillations of a star
and suppress stochastic effects.  The Q value is a measure of how many
oscillations there are before there is substantial damping of a mode.
Therefore, with the period of the oscillation mode scaling ($\nu_{\rm mode}$)
with the dynamical timescale, the effects of viscosity can be neglected if:
$$\nu_{\rm mode}=\left(\frac{R^3}{\sqrt{GM}}\right) \times Q > \sim P$$ with
$P$ being the orbital period.  For $P\sim 1$ year, this inequality is not
satisfied for main sequence stars with $Q < \sim10^6$. However, Piro (2011)
invokes higher $Q$-values for WDs in his work on WD binaries. He found
$Q=7\times 10^{10}$ and $Q = 2\times 10^7$, respectively for He and C/O
WDs. The above inequality is easily satisfied if we use $Q=7\times 10^{10}$,
while it is barely satisfied for $Q=2\times 10^7$. However, as stressed by
Piro (2011) the appropriate range of $Q$-values is still unknown for WDs. Some
authors have estimated Q-values up to $10^{12}$ (e.g. Campbell 1984) or
$10^{15}$ (Willems et al. 2010).

\subsection{Effects of particle resolution}
\label{resolution}

\begin{figure*}[ht!]
\includegraphics[scale=0.9]{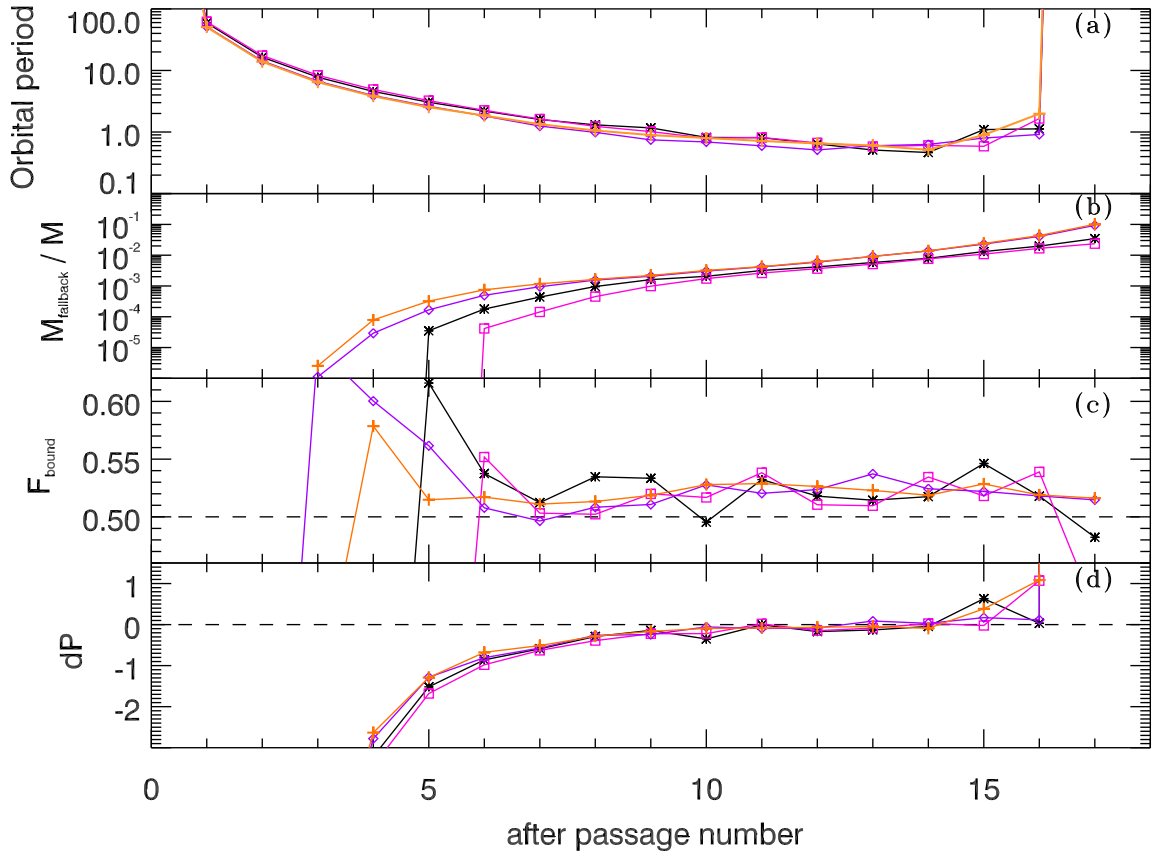}
\caption{Effects of the particle resolution $N$ for the case with $n =
  1.5$ \& $\Gamma = 5/3$ \& $r_p / r_t = 2.3$ on the evolution of the orbital
  period (a), the ratio of the fallback mass to the initial mass of the donor
  (b), the fraction of the stripped matter bound to the BH (c) and the change
  in the orbital period from one orbit to the next (d). In all cases, the
  inverse mass ratio $q^{-1} = 3\times 10^4$. The symbols correspond to
  squares for $N=2.5\times 10^4$, stars for $N=5\times 10^4$, diamonds for
  $N=10^5$ and crosses for $N=2\times 10^5$, respectively. The epoch when mass
  stripping starts happens earlier with higher particle resolution. Indeed,
  mass stripping starts at passage 6 for $N=2.5\times 10^4$, while for $N =
  1-2\times 10^5$ it starts at passage 3.
\label{fig:resolution}}
\end{figure*}

Fig.~\ref{fig:resolution} shows the effects of particle resolution ($N$) on
the evolution of the orbital period, the ratio of the fallback mass to the
initial mass of the donor, the fraction of the stripped matter bound to the BH
($F_{\rm bound}$) and the change in the orbital period from one orbit to the
next ($dP$) for the case with $r_p/r_t = 2.3$, $n=1.5$ \& $\Gamma=5/3$. In all
cases, the inverse mass ratio $q^{-1} = 3\times 10^4$.  Here we considered
$N$-values varying from $2.5\times 10^4$ to $2\times 10^5$ particles. From
this figure, we can see that overall the higher resolution simulations
behave qualitatively the same way as for lower resolution cases presented in
previous sections. Indeed, the donor is ejected from the system at passage 17
whatever the particle resolution used. We note that mass stripping appears to
begin sooner as $N$ increases, while $dP$ and $F_{\rm bound}$ 
appear to vary more smoothly at our highest resolution ($N=2\times 10^5$
particles). Even so, there still is a significant change in orbital period
between two successive passages in this latter case. So, we are confident that
the results discussed on this paper are robust.

\section{Conclusion}

After showing quasi-periodic outbursts in HLX-1 over the past four years, the
X-ray outburst in 2013 was delayed by more than a month. In this paper, we
investigated the implications of this delay in the framework of the mass
transfer model proposed by L11 and the orbital constraints derived from Webb
et al. (2014). To do so, we performed Newtonian SPH simulations over a large
parameter space for the donor type using polytropes with various equations of
state and the orbital parameters in order to study several points: 1) Could
the orbital parameters change suddenly in such a way to explain the delayed
outburst?; 2) Could an IMBH capture a star without disrupting it on a nearly
parabolic bound orbit? What will be the fate of such a system? In a
forthcoming paper, we will calculate the probability to form such highly
eccentric systems and estimate the space density of such systems in the local
Universe; 3) Could we put some constraints on the possible type of donor
stars?

Our results show that an incoming star approaching a BH on a parabolic orbit
could become bound to the BH with an orbital eccentricity close to 1. After
the formation of such an highly elliptical binary system, the orbital period
will decrease until reaching a minimum that can be shallow. Then, the orbital
period will tend to generally increase over several periapsis passages due to
tidal effects and increasing mass transfer. This will ultimately end with
the star being ejected. Depending on the reservoir of matter remaining in
the accretion disk, the BH will continue accreting this matter over a
timescale that could be much larger than the time during which the star was
bound to the BH. The source will then stay in the low/hard state and will
appear as a moderately bright ULX ($L_X \sim 10^{40}~{\rm erg~s}^{-1}$) until
switching to quiescence when running out of fuel.

In the cases matching the behavior of HLX-1, our simulations showed that the
donor star could perform several orbits before the system is disrupted in
qualitative agreement with our current observations. We also demonstrated the
importance of the effects of stochastic fluctuations inside the star that
appear to be tied to stochastic behavior in mass loss and that, by adding or
removing orbital energy from the system, could lead to sudden changes in the
orbital period from one periapsis passage to the next with the appropriate
order of magnitude of what has been observed for HLX-1.

Given the constraints on the BH mass ($M_{\rm BH} > 10^4~M_\odot$) and
assuming that the HLX-1 system is currently near a minimum in orbital period,
a recurrence time around 1 year implies that the donor star has to be a WD or
a stripped giant core passing at periapsis at only a few gravitational radii
from the IMBH. We also discussed the possibility that given the substantial
amount of matter transferred at periapsis this might give rise to an extreme
and brief supercritical accretion episode. Although spectral fitting of the
X-ray data do not favor supercritical accretion, we speculate that such high
accretion episodes might be missed by observations not done within a fraction
of a day preceding the start of the outburst. So, the increase in mass
transferred to the accretion disk at each periapsis is an important process to
consider in more detail and possibly the biggest challenge to the MTM presented
in this paper.

\acknowledgments We would like to thank J.-P. Lasota, G. Dubus \& A. King for
useful discussions. We thank the anonymous referee for his/her comments. 
  JCL is supported by National Science Foundation (NSF) Grant No. AST-1313091.
  This work used the Extreme Science and Engineering Discovery Environment
  (XSEDE), which is supported by NSF grant number OCI-1053575.

{\it Facilities:} \facility{Swift}.

\begin{deluxetable}{cccllccc}
\scriptsize
\tablecolumns{8} \tablewidth{0pc} \tablecaption{Response of orbital parameters and mass loss to multiple periapsis passages.
\label{sph_table}}
\tablehead{
\colhead{after passage}&\colhead{orbital} & \colhead{semimajor}& \colhead{}& \colhead{}& \colhead{peak fallback}\\
\colhead{number} & \colhead{period} & 
\colhead{axis $a/R$} & \colhead{eccentricity} & 
\colhead{$M_{\rm donor}/M$} & \colhead{$dM/dt$} & \colhead{$M_{\rm fallback}/M$} &\colhead{$t_{\rm Edd}$}
}
\startdata
\hline
&&&$ n=1.5$, $\Gamma=5/3$, $ r_{\rm p}=2.4 r_{\rm t}$\\
         1&  158.    & 0.135E+08& 0.99999    &  1.0000    &   0.0    &   0.0    &   0.0    \\
         2&  44.4    & 0.581E+07& 0.99999    &  1.0000    &   0.0    &   0.0    &   0.0    \\
         3&  20.5    & 0.347E+07& 0.99998    &  1.0000    &   0.0    &   0.0    &   0.0    \\
         4&  12.2    & 0.245E+07& 0.99997    &  1.0000    &   0.0    &   0.0    &   0.0    \\
         5&  8.13    & 0.187E+07& 0.99996    &  1.0000    &   0.0    &   0.0    &   0.0    \\
         6&  5.87    & 0.151E+07& 0.99995    &  1.0000    &   0.0    &   0.0    &   0.0    \\
         7&  4.40    & 0.124E+07& 0.99994    & 0.99999    &  0.17    &  0.33E-05&  0.20    \\
         8&  3.52    & 0.107E+07& 0.99993    & 0.99995    &  0.56    &  0.28E-04&  0.63    \\
         9&  2.85    & 0.932E+06& 0.99992    & 0.99981    &   1.7    &  0.77E-04&   1.1    \\
        10&  2.33    & 0.814E+06& 0.99991    & 0.99957    &   2.0    &  0.13E-03&   1.4    \\
        11&  2.11    & 0.761E+06& 0.99990    & 0.99909    &   4.3    &  0.28E-03&   2.1    \\
        12&  1.90    & 0.709E+06& 0.99989    & 0.99832    &   4.9    &  0.44E-03&   2.7    \\
        13&  1.64    & 0.644E+06& 0.99988    & 0.99714    &   7.0    &  0.64E-03&   3.3    \\
        14&  1.49    & 0.604E+06& 0.99988    & 0.99549    &  10.6    & 0.891E-03&   3.9    \\
        15&  1.37    & 0.572E+06& 0.99987    & 0.99334    &  13.8    & 0.113E-02&   4.5    \\
        16&  1.30    & 0.551E+06& 0.99986    & 0.99097    &  18.0    & 0.128E-02&   4.7    \\
        17&  1.23    & 0.531E+06& 0.99986    & 0.98801    &  20.6    & 0.156E-02&   5.3    \\
        18&  1.04    & 0.476E+06& 0.99984    & 0.98421    &  26.3    & 0.196E-02&   6.0    \\
        19& 0.898    & 0.431E+06& 0.99983    & 0.97979    &  29.1    & 0.226E-02&   6.4    \\
        20& 0.799    & 0.399E+06& 0.99981    & 0.97476    &  41.5    & 0.264E-02&   7.0    \\
        21& 0.792    & 0.396E+06& 0.99981    & 0.96800    &  54.5    & 0.358E-02&   8.2    \\
        22& 0.754    & 0.384E+06& 0.99981    & 0.95963    &  60.0    & 0.441E-02&   9.2    \\
        23& 0.808    & 0.402E+06& 0.99981    & 0.94907    &  84.3    & 0.552E-02&   10.    \\
        24& 0.834    & 0.410E+06& 0.99982    & 0.93418    &  125.    & 0.776E-02&   12.    \\
        25&  1.46    & 0.595E+06& 0.99987    & 0.91431    &  175.    & 0.106E-01&   15.    \\
        26&  1.19    & 0.521E+06& 0.99986    & 0.88596    &  220.    & 0.147E-01&   17.    \\
        27&  26.9    & 0.416E+07& 0.99998    & 0.84008    &  350.    & 0.243E-01&   23.    \\
        28& Infinity &-0.370E+07&  1.0000    & 0.75818    &  544.    & 0.424E-01&   30.    \\
\hline
&&&$ n=2$, $\Gamma=1.5$, $ r_{\rm p}=2.3 r_{\rm t}$\\
        26&  1.68    & 0.653E+06& 0.99989    & 0.97546    &  25.5    & 0.183E-02&   5.7    \\
        27&  1.60    & 0.633E+06& 0.99989    & 0.97110    &  34.4    & 0.231E-02&   6.5    \\
        28&  1.71    & 0.662E+06& 0.99989    & 0.96559    &  43.2    & 0.296E-02&   7.4    \\
        29&  1.56    & 0.623E+06& 0.99989    & 0.95839    &  54.0    & 0.372E-02&   8.4    \\
        30&  1.87    & 0.703E+06& 0.99990    & 0.94886    &  71.2    & 0.502E-02&   9.8    \\
        31&  2.84    & 0.928E+06& 0.99992    & 0.93506    &  103.    & 0.738E-02&   12.    \\
\hline
\tablebreak
&&&$ n=2$, $\Gamma=5/3$, $ r_{\rm p}=2.3 r_{\rm t}$\\
        50&  1.12    & 0.498E+06& 0.99986    & 0.95619    &  23.8    & 0.136E-02&   4.9    \\
        51&  1.22    & 0.529E+06& 0.99986    & 0.95378    &  23.8    & 0.132E-02&   4.8    \\
        52&  1.12    & 0.499E+06& 0.99986    & 0.95099    &  24.5    & 0.143E-02&   5.0    \\
        53&  1.18    & 0.517E+06& 0.99986    & 0.94800    &  32.9    & 0.160E-02&   5.4    \\
        54&  1.37    & 0.571E+06& 0.99987    & 0.94486    &  37.9    & 0.169E-02&   5.5    \\
        55&  1.20    & 0.523E+06& 0.99986    & 0.94153    &  31.6    & 0.171E-02&   5.5    \\
\hline
&&&$ n=2$, $\Gamma=5/3$, $ r_{\rm p}=2.4 r_{\rm t}$\\
        56&  1.40    & 0.581E+06& 0.99987    & 0.98405    &   4.6    &  0.29E-03&   2.2    \\
        57&  1.32    & 0.557E+06& 0.99987    & 0.98341    &   5.1    &  0.30E-03&   2.2    \\
        58&  1.39    & 0.576E+06& 0.99987    & 0.98282    &   5.7    &  0.35E-03&   2.4    \\
        59&  1.40    & 0.580E+06& 0.99987    & 0.98219    &   6.6    &  0.35E-03&   2.4    \\
        60&  1.36    & 0.568E+06& 0.99987    & 0.98145    &   6.4    &  0.38E-03&   2.5    \\
        61&  1.32    & 0.557E+06& 0.99987    & 0.98076    &   6.9    &  0.35E-03&   2.4    \\
        62&  1.32    & 0.557E+06& 0.99987    & 0.98006    &   6.1    &  0.38E-03&   2.5    \\
        63&  1.51    & 0.608E+06& 0.99988    & 0.97928    &   8.1    &  0.46E-03&   2.8    \\
        64&  1.46    & 0.597E+06& 0.99988    & 0.97848    &   7.6    &  0.40E-03&   2.6    \\
...\\
        68&  1.21    & 0.527E+06& 0.99986    & 0.97521    &   8.0    &  0.37E-03&   2.5    \\
        69&  1.29    & 0.549E+06& 0.99986    & 0.97437    &   8.8    &  0.48E-03&   2.8    \\
        70&  1.23    & 0.532E+06& 0.99986    & 0.97350    &   6.4    &  0.45E-03&   2.7    \\
        71&  1.32    & 0.557E+06& 0.99987    & 0.97243    &   8.9    &  0.61E-03&   3.2    \\
        72&  1.49    & 0.604E+06& 0.99988    & 0.97160    &   8.2    &  0.51E-03&   2.9    \\
        73&  1.55    & 0.620E+06& 0.99988    & 0.97052    &   9.6    &  0.61E-03&   3.2    \\
\hline
&&&$ n=2$, $\Gamma=5/3$, $ r_{\rm p}=2.5 r_{\rm t}$\\
       143& 0.671    & 0.355E+06& 0.99978    & 0.95006    &  13.1    & 0.621E-03&   3.2    \\
       144& 0.658    & 0.350E+06& 0.99978    & 0.94879    &  11.4    & 0.666E-03&   3.4    \\
       145& 0.670    & 0.354E+06& 0.99978    & 0.94769    &  13.2    & 0.580E-03&   3.1    \\
       146& 0.717    & 0.371E+06& 0.99979    & 0.94640    &  13.3    & 0.747E-03&   3.6    \\
       147& 0.809    & 0.402E+06& 0.99981    & 0.94507    &  16.4    & 0.778E-03&   3.7    \\
       148& 0.839    & 0.412E+06& 0.99981    & 0.94369    &  13.3    & 0.741E-03&   3.6    \\
\hline
&&&$ n=2$, $\Gamma=5/3$, $ r_{\rm p}=2.7 r_{\rm t}$\\
       317&  1.63    & 0.640E+06& 0.99987    & 0.95578    &   3.8    &  0.22E-03&   1.9    \\
       318&  1.60    & 0.633E+06& 0.99987    & 0.95521    &   5.5    &  0.29E-03&   2.2    \\
       319&  1.69    & 0.656E+06& 0.99987    & 0.95465    &   5.5    &  0.33E-03&   2.3    \\
       320&  1.71    & 0.663E+06& 0.99987    & 0.95401    &   7.5    &  0.37E-03&   2.5    \\
       321&  1.88    & 0.706E+06& 0.99988    & 0.95343    &   5.8    &  0.36E-03&   2.4    \\
       322&  1.86    & 0.700E+06& 0.99988    & 0.95272    &   7.5    &  0.36E-03&   2.4    \\
...\\
       328&  1.89    & 0.707E+06& 0.99988    & 0.94854    &   6.3    &  0.40E-03&   2.6    \\
       329&  1.91    & 0.713E+06& 0.99988    & 0.94778    &   4.9    &  0.40E-03&   2.6    \\
       330&  1.99    & 0.734E+06& 0.99989    & 0.94703    &   7.4    &  0.43E-03&   2.7    \\
       331&  2.01    & 0.738E+06& 0.99989    & 0.94633    &   8.0    &  0.38E-03&   2.5    \\
       332&  2.40    & 0.829E+06& 0.99990    & 0.94533    &   9.2    &  0.57E-03&   3.1    \\
       333&  2.18    & 0.779E+06& 0.99989    & 0.94448    &   8.1    &  0.43E-03&   2.7    \\
\enddata
\tablenotetext{a}{The time unit for the orbital period is $t_{\rm u}\equiv {\rm yr}\,\lambda^{3/2}M_4^{-1/2}$, where $\lambda\equiv R/(0.01R_\odot)$ and $M_4\equiv M_{\rm BH}/(10^4 M_\odot)$.  The unit of fallback $dM/dt$ is $M/t_u=M_\odot {\rm yr}^{-1}\mu\lambda^{-3/2}M_4^{1/2}$, where $\mu=M/M_\odot$.
The Eddington time $t_{\rm Edd}$ has been calculated according to eq.\ (\ref{tEdd}) and is listed in units of ${\rm day}\,(\epsilon/0.1)^{0.53}\mu^{0.53}\lambda^{0.71}M_4^{-0.76}$, where $\epsilon$ is the radiative efficiency.
}
\end{deluxetable}

\end{document}